\documentclass[aps,pra,showpacs,twoside,twocolumn,10pt]{revtex4-1}
\usepackage[colorlinks=true, citecolor=red, urlcolor=blue ]{hyperref}
\usepackage{epsfig,newlfont,amssymb,amsfonts,amsmath,bm,subfigure,palatino,mathtools,amsthm,braket,times,soul,enumitem,color}
\usepackage[normalem]{ulem}

\usepackage{blindtext}
\usepackage{graphicx}
\usepackage{amsmath}
\usepackage{bm}
\usepackage{hyperref}
\usepackage{geometry}
\usepackage{amsthm}
\usepackage{gensymb}
\usepackage{physics}
\begin{document}

\title{Quantum analog of resource theory of stinginess}
\author{Ujjwal Sen}
\affiliation{Harish-Chandra Research Institute, HBNI, Chhatnag Road, Jhunsi, Allahabad 211 019, India}
\date{01 April 2021}

\begin{abstract}
   We present a resource theory of stinginess, first in a classical scenario, and then, point to a possible quantum version  of the same. 
\end{abstract}
\maketitle
%\section{Introduction}

Resource theories play an important role in the rapidly developing field of quantum information~\cite{Nielsen-Chuang;Preskill;Bruss-Leung}, and was the main motivation behind this work, even though such theories possibly pervade all scientific streams. 
Within the field of quantum information, the resource theory of entanglement~\cite{Horodecki;Sreetama-Titas} is probably the most popular and arguably the most well-understood and useful, although several other resource theories have been formalized in the past decade or so, including those related to quantum coherence~\cite{coherence}. 
Here we present a quantum theory of stinginess, beginning first with a classical set-up.

We were led to this resource theory by the following question and answer that we found on social media.\\ 

\noindent \textbf{Question:} A person, named Scrooge~\cite{Scrooge}, who is well-known to be thrifty, accidentally breaks a tooth of a comb.
%~\cite{comb}. 
He forthwith goes to a shop and buys a new comb. How do you explain the person's behavior, given that he is a miser?
%~\cite{feminism}
\\

\noindent \textbf{Answer:} Because that was the last tooth of the comb. [He of course did not possess any other comb!]\\

Inspired by this question and answer, we can set up a classical resource theory of stinginess. Suppose that a person possesses  a comb with \(n\) teeth (along with a shaft), with \(n\) being a (finite) nonnegative number. 
%This is in possession of a person. 
The person will be called maximally stingy if s/he buys a comb only when \(n\) teeth break off from the comb. Scrooge is an example of a maximally stingy person. The amount of stinginess, denoted by \(S_C\), of a certain person in possession of a comb with \(n\) teeth, can  therefore take a discrete and finite set of values, viz. the integers \(0\) through \(n\), both included, divided by \(n\). For Scrooge, we have 
\begin{equation}
S_C(\text{Scrooge})=1.
\end{equation}
The same is true for any maximally stingy person. Unlike many resource theories in quantum information, the measure takes only a discrete set of values. Having a discrete set of values for a resource is of course nothing new, and indeed, all resources in classical computer science and many resources in its quantum cousin  
%computer science 
are discrete in nature. 
%Having an application is of course a hallmark of a successful resource theory, and the current one can, e.g., be useful to identify the people to be referred to a future climate change tribunal, as it could possibly be argued that the said group of people will have a close resemblance with the set of people for whom \(S_C=0\). 

In the quantum case, we can model the comb as a quantum register having \(n\) qubits, with the laboratory playing the role of the shaft and the qubits of the teeth. The laboratory is ``manned'' by a woman named Chippu. Her stinginess is to be measured with respect to the quantum comb that she runs in her laboratory. The quantum comb is used for some task that in the optimal case requires
\(n\) qubits. However, the task does run, albeit non-optimally, with less than \(n\) qubits also. We also assume that running the device optimally for the said task requires the \(n\) qubits to be in an arbitrary state of a particular (``preferred'') product orthonormal basis of the \(n\)-qubit Hilbert space \((\mathbb{C}^2)^{\otimes n}\). This is a typical requirement for the initial state of the circuit in a quantum algorithm~\cite{Nielsen-Chuang;Preskill;Bruss-Leung}. Therefore, a natural measure of quantum stinginess of Chippu and her laboratory will be a function of  \(m\), the number of qubits lost, and the initial state of the \(n\)-qubit quantum comb before losing any qubit. Of course, \(0\leq m \leq n\). Let us denote this quantity as \(S_Q(m,\varrho_n)\), with \(\varrho_n\) being an \(n\)-qubit quantum state. If \(S_Q(m,\varrho_n)\) is higher than a previously-decided value \(S_Q^0\), for a particular \(m\) and \(\varrho_n\), Chippu resets the quantum comb. One way of defining the quantity is by measuring the minimal distance, to the elements of the preferred product basis, of the state \(\varrho_n\) after \(m\) arbitrary qubits are removed.
%, and replaced by arbitrary auxiliaries. 
More formally, therefore, 
\begin{equation}
S_Q(m,\varrho_n) = 
%\phantom{tiler-naRu-kheye-nanda-nachi}\nonumber\\
\frac{1}{2}\left[\frac{m}{n}+\min_{i, \text{tr}_m}
\mathcal{D}\left( \text{tr}_m \varrho_n , \text{tr}_m \left(|\psi_i\rangle \langle \psi_i|\right)  \right)\right],
%\nonumber\\
\end{equation}
for \(m<n\). For \(m=n\), we set \(S_Q(m,\varrho_n) = \infty\) for all \(\varrho_n\). The minimization over \(\text{tr}_m\) indicates that any \(m\) of the \(n\) qubits could be lost, and we consider the stingy-case scenario to define the measure. Here, \(\mathcal{D}\) is a measure of distance, normalized to unity, on the space of density operators on \((\mathbb{C}^2)^{\otimes n}\), and \(\{|\psi_i\rangle\}_{i=1}^{2^n}\) is the preferred product basis on \((\mathbb{C}^2)^{\otimes n}\).
Using distance-based concepts has a long history in quantum information, and in particular, one remembers the conceptualization of  relative entropy of entanglement~\cite{relent}. Taking a distance to product states has also been used in several measures. In particular, see Refs. \cite{ Grover+, wd, GM, sharedpurity}.

Along with a measure to quantify the resource, a resource theory usually contains a set of free states and a set of free operations. For a given value of \(S_Q^0\) and \(m\), the set of free states of the \(n\)-qubit quantum comb of Chippu is given by those \(\varrho_n\) for which 
\begin{equation}
S_Q(m, \varrho_n) \leq S_Q^0.
\end{equation} 
Let us denote this set by \(\mathcal{F}^m_{S_Q^0}\). The set of free operations, on the other hand, is given by those quantum operations for which an arbitrary state of \(\mathcal{F}^m_{S_Q^0}\) will necessarily be taken to one of the same set. 

%Just like for classical resource theories, 
An application is of course a trademark of a satisfactory resource theory, whether in the classical or in the quantum domain. We identify two such applications of the resource theory of stinginess. The \emph{second} one is that the theory can be useful in identifying people to be referred to a future climate change tribunal, as it could possibly be argued that the said group of people will have a close resemblance with the set of people for whom \(S_C=0\) or  who sets a low \(S_Q^0\). 

\begin{acknowledgments}
The question and answer, and the classical resource theory of stinginess, came up during a discussion with Sohail and Saronath Halder. We also acknowledge a discussion some 30 years ago with Shanti Ram Ghosh.
We were unable to identify the  person who came up with the question and answer with which we began the manuscript. We acknowledge support from the Department of Science and Technology, Government of India through the QuEST  grant (grant number DST/ICPS/QUST/Theme-3/2019/120).
\end{acknowledgments}

\end{document}